\documentclass[twocolumn,pra,showpacs,preprintnumbers,amsmath,amssymb,floatfix]{revtex4}
\usepackage{graphicx}
\usepackage{dcolumn}
\usepackage{bm}

\begin{document}
\DeclareGraphicsExtensions{.eps}

\title{Density Matrix Tomography of Entangled Electron and Nuclear Spin States in $\mathrm{^{15}N@C_{60}}$}
\author{Werner Scherer}
\author{Michael Mehring}
\email{m.mehring@physik.uni-stuttgart.de}
\affiliation{2. Physikalisches Institut, Universit\"at Stuttgart,
Pfaffenwaldring 57, 70550 Stuttgart, Germany}

\date{\today}

\begin{abstract}
We discuss details of the
preparation and detection of entangled electron-nuclear spin
states in $\mathrm{^{15}N@C_{60}}$ together with a quantitative
evaluation of the complete density matrix. All four Bell states
of a two qubit subsystem were analyzed. In addition we find a
quantum critical temperature of $T_q = 7.73~K$ for this system at
an electron spin resonance frequency of $95~GHz$.
\end{abstract}

\pacs{03.67.Mn, 03.65.Ud, 33.40.+f, 72.80.Rj}
 \maketitle

\section{Introduction}
    In recent years quantum information theory has come to exciting new ideas in the fields of quantum teleportation \cite{bennett:93,bouwmeester:97},
 quantum cryptography \cite{bennett:92} and quantum computation \cite{deutsch:85,deutsch:92}. It was stimulated enormously by the discovery of the
factoring \cite{shor:97,vandersypen:01} and searching algorithms
\cite{grover:97,chuang:98b}, which demonstrate that a quantum
computer is capable of tasks, that are impossible to solve on a
classical computer. Consequently, over the past few years various
physical systems were examined or proposed for their potential
use as hardware for quantum computing. These include ions
confined to an electromagnetic trap \cite{cirac:95,monroe:95},
photons \cite{knill:01}, doped semiconductors \cite{kane:98},
Cooper pair states in superconductors
\cite{shnirman:97,nakamura:99}, Josephson junctions
\cite{mooij:99,friedmann:00}, and nuclear spins in liquids (NMR
quantum computing)
\cite{cory:97,cory:98a,gershenfeld:97a,knill:98a,jones:98,warren:97,mehring:99}.

One of the most interesting components in many quantum algorithms
are nonlocal quantum correlations that violate our conception of
the classical world. These so called entangled states have been
discussed for a long time
\cite{einstein:35,bell:64,greenberger:89,mermin:90,lloyd:98}.
Experimental preparations of entangled states of photons have
been published\cite{zeilinger:97, tittel:98a}. The pseudo
entanglement of three coupled nuclear spins in the so-called
Greenberger-Horne-Zeilinger (GHZ) \cite{greenberger:90} state was
demonstrated by
NMR\cite{laflamme:98,nelson:00,teklemariam:01,teklemariam:02}.
Entanglement was also achieved among ions confined in an
electromagnetic trap \cite{sackett:00}.

In this contribution we present procedures for creating and
detecting entanglement between electronic and nuclear spin states
in the solid phase of the endohedral fullerene
$\mathrm{^{14}N@C_{60}}$ which has already been proposed as a
basic unit for a scalable quantum computer
\cite{harneit:02,suter:02}. We will show, however, that
$\mathrm{^{15}N@C_{60}}$ is not a qubit but rather a multi-qubit
system and specific addressing schemes must be applied when
considering all quantum levels. From its multiple quantum states
we will project to a two qubit subsystem and demonstrate the
generation of entanglement within this subsystem. The
experimental approach is related to the preliminary work
presented in \cite{mehring:03} where entanglement has been
simulated between an electron spin $1/2$ and a nuclear spin $1/2$
of a radical electron in a single crystal of malonic. Due to the
strong hyperfine interaction in the system presented here
entanglement can be achieved by a factor of thousand shorter time
scale than in liquid state NMR. A brief account of our approach
to $\mathrm{^{14}N@C_{60}}$ was already published in
\cite{mehring:04a}. Here we present further details on the
preparation and tomography.

The experiments reported here are carried out in a mixed spin
state of an ensemble of spins. Since the complete density matrix
in such a system is {\em separable} strictly speaking
entanglement like in pure quantum states can not be achieved
\cite{braunstein:99, schack:99}. However, the unitary operations
applied to the mixed states density matrices are identical to
those applied to pure systems and moreover, the resulting density
matrices have the same operator structure as the ones of the pure
system except for a different overall factor. Therefore these
states are called pseudo pure or pseudo entangled
\cite{cory:97,gershenfeld:97a,knill:98a}. In the end we will show
that these restrictions can be overcome for the system under
investigation for high enough magnetic fields ($95$~GHz electron
spin resonance frequency)and low enough temperature( $T =
7.73~K$) with current technology. The quantum limit where pseudo
entangled states become entangled states is therefore well in
reach.

Before we start to present the procedures for the more
complicated case of $\mathrm{^{15}N@C_{60}}$ we sketch as a
reminder how entanglement could be achieved in a system of two
coupled quantum bits (qubits) represented for example by an
electron  spin $S=1/2$ and a nuclear spin I=1/2.
    Starting from the Zeeman product states
    \begin{equation}
        |m_S\,m_I\rangle=|\uparrow \uparrow\rangle, \, |\uparrow \downarrow\rangle, \,
            |\downarrow \uparrow\rangle, \, |\downarrow \downarrow\rangle\,,
    \end{equation}
entanglement is achieved by applying first a Hadamard
transformation $H$ on one spin followed by a controlled not
(CNOT) operation. Suppose we start from the pure state
$|\downarrow\uparrow\rangle$ we can create the entangled state
$\Psi^+$ by
    \begin{equation}
    |\downarrow\uparrow\rangle\stackrel{\mbox{H}}{\rightarrow} \frac{1}{\sqrt{2}}|\downarrow\downarrow\rangle+
    |\downarrow\uparrow\rangle\stackrel{\mbox{CNOT}}{\longrightarrow} \frac{1}{\sqrt{2}}|\uparrow\downarrow\rangle+
    |\downarrow\uparrow\rangle\, .
    \end{equation}
    Depending on the applied unitary transformations and initial states,
    all four entangled states (Bell-basis) of a two qubit system can be generated by this procedure:
    \begin{equation}
        \left| \Phi^{\pm} \right\rangle=
        \frac{1}{2}\left(\left|\uparrow \uparrow \right\rangle\pm\left|\downarrow \downarrow \right\rangle\right)\,
        ,\quad
        \left| \Psi^{\pm} \right\rangle=
        \frac{1}{2}\left(\left|\uparrow \downarrow \right\rangle\pm\left|\downarrow \uparrow
        \right\rangle\right)\, .\label{bell}
    \end{equation}
    As we will see later in detail those transformations
    can be realized by selective high frequency pulses applied to allowed transitions of
    the system. We will use in the following the qubit notation
    $|m_S\,m_I\rangle=|00\rangle, \, |01\rangle, \,|10\rangle, \,
    |11\rangle$ as well as the state labelling
    $|1\rangle, \, |2\rangle, \,|3\rangle, \,
    |4\rangle$.

\subsection{Endohedral fullerene  $\mathbf{\mathrm{^{15}N@C_{60}}}$}

    \begin{figure}[htb]
     \centerline{\includegraphics[width=0.3\textwidth]{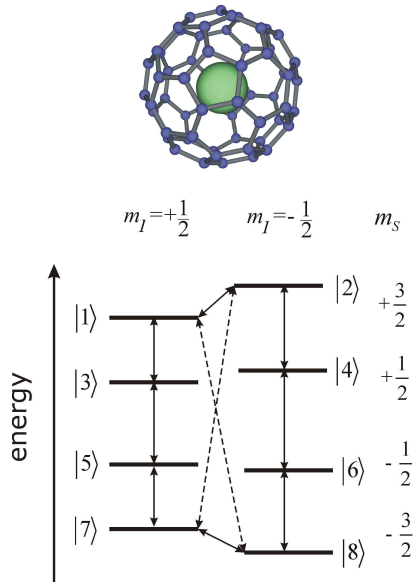}}
        \caption{The eight quantum states of a two spin system with $S=3/2$ and $I=1/2$. The
          arrows denote allowed transitions. The dotted arrows indicate
          forbidden transitions that belong to the entangled states discussed in this publication.
          The separation of the energy levels is not to scale.}\label{8niveau}
    \end{figure}

The experiments were performed on a powder sample of
$\mathrm{^{15}N@C_{60}}$. The $^{15}\mathrm{N}$ atom resides in
the center of the $\mathrm{C_{60}}$ molecule. There is no charge
transfer from the atom to the cage. The nitrogen atom is
paramagnetic due to the half filled $p$ orbital shell with three
unpaired electrons which form a total electron spin $S=3/2$.
Because of the $\mathrm{^{15}N}$ isotope enrichment the nuclear
spin of the nitrogen atom is $I=1/2$.  The magnetic resonance
properties of the sample have already been discussed in detail in
\cite{almeidamurphy:96,weiden:99}. The nitrogen atoms were
implanted into the carbon cage by Weidinger and
co-workers\cite{pietzak:97} by simultaneous evaporation of
$\mathrm{C_{60}}$ and ion bombardment onto a cooled target. The
sample was purified by high-pressure liquid chromatography.

With the magnetic field oriented along the $z$-axis, the
Hamiltonian of this system becomes
    \begin{equation}
    \mathcal{H} =\hbar\left(\omega_S S_z+\omega_I I_z+2\pi a\mathbf{S}\mathbf{I}\right)\,.
    \end{equation}
where $\omega_{S}$ and $\omega_ {I}$ are the Larmor frequencies
of electron spin $S=3/2$ and the nuclear spin $I=1/2$. Due to the
high symmetry of the molecule the hyperfine interaction
$a=-22.08$~MHz is isotropic. This value is larger than for a free
nitrogen \cite{anderson:59} atom because of the confinement
inside the carbon cage. For strong enough magnetic fields, where
$|a|,\,\omega_I\ll\omega_S$, second order contributions to the
hyperfine coupling can be neglected and the Hamiltonian is in
first order given by:

\begin{equation}
        \mathcal{H}=\hbar\left(\omega_{S}S_z+\omega_{I}I_z+2\pi
        a\,S_z\,I_z\right)\,.
\end{equation}

The eight eigenvalues of this system are

    \begin{equation}
        E_{m_S\,m_I}=\hbar\left(\omega_{S} m_S+\omega_{I} m_I + 2\pi a\, m_S
        \,m_I\right)\, ,
    \end{equation}

depending on the spin quantum numbers $m_S=\pm 1/2$, $\pm3/2$ of
the electron spin and $m_I=\pm 1/2$ of the nuclear spin. The
eight eigenstates of the system are given by
    \begin{eqnarray}
        \left|m_S\,m_I\right\rangle&=&\left|1\right\rangle,\left|2\right\rangle,\ldots,\left|8\right\rangle\\
        &=&\left|+\frac{3}{2}\,+\frac{1}{2}\right\rangle,\left|+\frac{3}{2}\,-\frac{1}{2}\right\rangle,\ldots,\nonumber
        \left|-\frac{3}{2}\,-\frac{1}{2}\right\rangle\, .
    \end{eqnarray}

\begin{figure}[htb]
      \centerline{\includegraphics[width=0.4\textwidth]{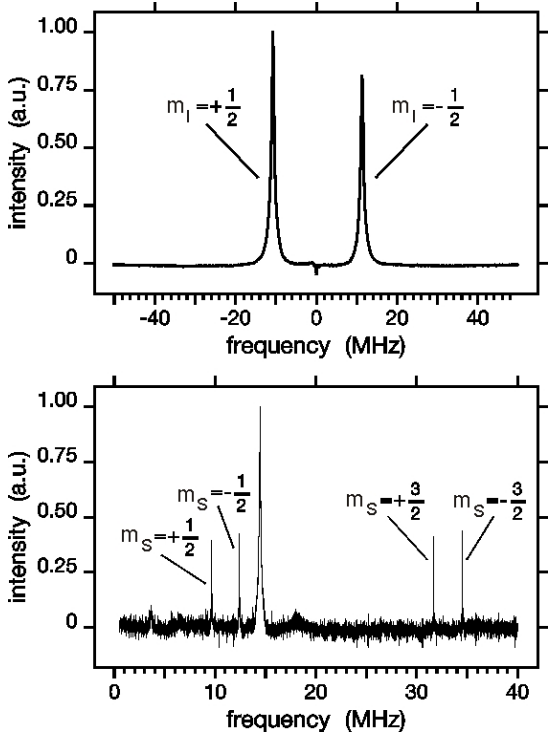}}
        \caption[ESR and ENDOR spectrum of $\mathrm{^{15}N@C_{60}}$]{ X-band ESR (top) and pulsed-ENDOR
        spectra \cite{mims:65}(bottom) of a powder sample of $\mathrm{^{15}N@C_{60}}$ in a magnetic field of
        $340$~mT. The ESR spectrum was obtained by Fourier transformation of the electron free
        induction decay (FID).}\label{15nc60spec}
    \end{figure}

The corresponding energy level scheme is shown in Fig.
\ref{8niveau}. Allowed transitions are indicated by arrows. There
are in first order three degenerate electron spin transitions
($\Delta m_S=\pm1$) for $m_I=+1/2$ and and another three for
$m_I=-1/2$. Because of this degeneracy the transitions can only be
excited simultaneously. There are, however, four nuclear spin
transitions ($\Delta m_I=\pm1$). The resulting electron spin
resonance (ESR) and electron nuclear double resonance (ENDOR)
spectra shown in Fig. \ref{15nc60spec} verify this level scheme.
The lines are labelled according to the spin states $m_I$ and
$m_S$. The ESR spectrum exhibits two lines separated by
$22.08$~MHz. The ENDOR spectrum consists of four lines located at
$9.67$~MHz, $12.41$~MHz, $31.69$~MHz and $34.54$~MHz.

\subsection{Entangled spin states in $\mathrm{^{15}N@C_{60}}$}

Instead of the four entangled Bell states for two coupled spins
1/2 the endohedral fullerene $\mathrm{^{15}N@C_{60}}$ allows for
$2\times12$ entangled states of the type
    \begin{equation}
        \Psi^{\pm}_{jk}=\frac{1}{\sqrt{2}}\left(\left|j\right\rangle\pm\left|k\right\rangle\right)
    \end{equation}

with $jk\in\left\{14,23,36,45,58,67,16,25,38,47,18,27\right\}$.
Here we will restrict ourselves to the preparation and detection
of the following two examples

\begin{eqnarray}
\left|\Psi_{27}^{\pm}\right\rangle&=&\frac{1}{\sqrt{2}}\left(\left|2\right\rangle\pm\left|7\right\rangle\right)\label{psi27}\\
&=&\nonumber            \frac{1}{\sqrt{2}}\left(\left|+\frac{3}{2}\,-\frac{1}{2}\right\rangle\pm\left|-\frac{3}{2}\,+\frac{1}{2}\right\rangle\right)
\\&\equiv&\nonumber
\frac{1}{\sqrt{2}}\left(\left|01\right\rangle\pm\left|10\right\rangle\right)\\
\left|\Phi_{18}^{\pm}\right\rangle&=&\frac{1}{\sqrt{2}}\left(\left|1\right\rangle\pm\left|8\right\rangle\right)\label{phi18}\\
&=&\nonumber            \frac{1}{\sqrt{2}}\left(\left|+\frac{3}{2}\,+\frac{1}{2}\right\rangle\pm\left|-\frac{3}{2}\,-\frac{1}{2}\right\rangle\right)\\
&\equiv&\nonumber
\frac{1}{\sqrt{2}}\left(\left|00\right\rangle\pm\left|11\right\rangle\right)
\end{eqnarray}

Except for the electron spin state with $m_S=\pm3/2$ these states
are equivalent to the two qubit Bell states of Eq. (\ref{bell}).
The corresponding coherent superpositions are indicated in Fig.
\ref{8niveau} by dashed arrows. This defines the fictitious two
state subsystem of the electron spin with $|\pm3/2\rangle$ as one
qubit. For the second qubit we consider the $|\pm1/2\rangle$
states of the nuclear spin. In other words the quantum states
$|\pm3/2\,\pm1/2\rangle$ represent the four states of our two
qubit subsystem. Of course there are other combinations of
quantum states that could be regarded as an independent two
qubit subsystem. In this sense  the quantum system of
$\mathrm{^{15}N@C_{60}}$ represents a multi qubit system. In the
following we will use the qubit notation  $|00\rangle$,
$|01\rangle$, $|10\rangle$ and $|11\rangle$ for the states
$|+\frac{3}{2}\,+\frac{1}{2}\rangle$,
$|+\frac{3}{2}\,-\frac{1}{2}\rangle$,
$|-\frac{3}{2}\,+\frac{1}{2}\rangle$ and
$|-\frac{3}{2}\,-\frac{1}{2}\rangle$.

\subsection{{\large$\mathbf{z}$}-rotations of the quantum states\label{z-rot}}
In the tomography of quantum states we will apply extensively
phase rotations about the $z$-axis. Therefore we investigate here
briefly how the different quantum states behave under these phase
rotations. A rotation of a spin state of spin $S$ by angle
$\varphi_1$ and spin $I$ by phase $\varphi_2$ around the
quantization axis ($z$-axis),
    corresponds to the unitary transformations
    \begin{equation}
        U_{S_z}=e^{-i\varphi_1S_z}\quad\mbox{and}\quad
    U_{I_z}=e^{-i\varphi_2I_z}\,.
    \end{equation}
    Under these transformations the quantum state $|m_S\,m_I\rangle$ exhibits the following phase
    variation:
    \begin{multline}
        U_{S_z}\,U_{I_z}|m_S\,m_I\rangle
    =e^{-i\varphi_1S_z}e^{-i\varphi_2I_z}|m_S\,m_I\rangle\\
    =e^{-i\left(m_S\varphi_1+m_I\varphi_2\right)}|m_S\,m_I\rangle
    \end{multline}

Let us apply these transformations to a superposition state of qubit 1 (subsystem with $m_S=\pm3/2$) which results in
    \begin{multline}
        \left(|00\rangle+|10\rangle\right)\left(\langle00|+\langle10|\right)
        \stackrel{z-\mbox{rot.}}{\longrightarrow}|00\rangle\langle00|+|10\rangle\langle10|\\
        +e^{-i 3\varphi_1}|00\rangle\langle10|
        +e^{i3\varphi_1}|10\rangle\langle00|.
      \label{nu1}
    \end{multline}
Note that in the case of an electron spin $S=1/2$ the factor $3$
in front of $\varphi_1$ should be replaced by $1$
\cite{mehring:03}. Consequently the superposition state of
nuclear spin states with $m_I=\pm1/2$ transforms like
    \begin{multline}
        \left(|10\rangle+|11\rangle\right)\left(\langle10|+\langle11|\right)
        \stackrel{z-\mbox{rot.}}{\longrightarrow}|10\rangle\langle10|+|11\rangle\langle11|\\
        +e^{-i\varphi_2}|10\rangle\langle11|+e^{i\varphi_2}|11\rangle\langle10|.
      \label{nu2}
    \end{multline}
    \vspace{2mm}
    Note, however, that applying phase rotations to entangled states
leads to a different behavior as is demonstrated here for the
$\Psi_{27}^{+}$ and $\Phi_{18}^{+}$ states:
\begin{subequations}
    \begin{multline}\label{eqn:nu1-nu2}
        \left(|01\rangle+|10\rangle\right)\left(\langle01|+\langle10|\right)
        \stackrel{z-\mbox{rot.}}{\longrightarrow}|01\rangle\langle01|+
        |10\rangle\langle10|\\
         + e^{-i(3\varphi_1-\varphi_2)}|01\rangle\langle10|+
          e^{i(3\varphi_1-\varphi_2)}|10\rangle\langle01|
    \end{multline}
    \begin{multline}\label{nu1+nu2}
        \left(|00\rangle+|11\rangle\right)\left(\langle00|+\langle11|\right)
        \stackrel{z-\mbox{rot.}}{\longrightarrow}|00\rangle\langle00|+|11\rangle\langle11|\\
        + e^{-i(3\varphi_1+\varphi_2)}|00\rangle\langle11|+
          e^{i(3\varphi_1+\varphi2)}|11\rangle\langle00|.
    \end{multline}
\end{subequations}
The entanglement is evidenced by the combined phase dependence
where both phase angles appear as a sum or difference depending
on the type of entangled state. In contrast the superposition
states of individual spins show only their corresponding single
phase behavior. This will be utilized to distinguish different
quantum states in the following.

\section{Experimental}\label{sec:experimental}
The experiments were performed on  a diluted powder sample of
$\mathrm{^{15}N@C_{60}}$ with a home-built X-band pulsed
spectrometer including a slotted tube resonator at about 9.5 GHz.
A radio frequency (rf) coil for the excitation of the ENDOR
transitions was inserted in the homogeneous region of the
microwave field with frequencies ranging from $0 - 40 $ MHz.
Both, the microwave channel at X-band frequencies as well as the
rf-channel were equipped with arbitrary waveform generators (AWG)
which allowed for arbitrary pulse and phase modulation. The
quantum states were controlled  by applying special unitary
transformations implemented as transition selective microwave and
radio frequency pulses with rotation angle $\beta$. These pulses
are labelled $P_{x;y}^{(\pm)}(\beta)$ for the two electron spin
transitions corresponding to $m_I=\pm1/2$ and
 $P_{x;y}^{(jk)}(\beta)$ for radio frequency pulses on nuclear spin transition
between levels $j$ and $k$ (see Fig. \ref{8niveau}). The
subscripts $x$ or $y$ indicate the phases of the pulses in the
corresponding rotating frame. The selective pulses can be written
in terms of fictitious spin $1/2$ operators $F_{x;y}^{(\pm)}$ or
$F_{x;y}^{(jk)}$, which belong to the transition $(\pm)$ or
$(jk)$ \cite{vega:77,vega:78} as
    \begin{equation}
       P_{x;y}^{\pm}(\beta)=\exp(-i\beta\,F_{x;y}^{\pm})\quad P_{x;y}^{jk}(\beta)=\exp(-i\beta\,F_{x;y}^{jk})\,.
    \end{equation}

   Pulses with arbitrary phase are expressed as

\begin{eqnarray}
  P_{x,\phi_1}^{\pm}(\beta)&=&e^{-i\phi_1F_{z}^{\pm}}e^{-i\beta\,F_{x}^{\pm}}e^{i\phi_1F_{z}^{\pm}}\quad \\
  P_{x,\phi_2}^{jk}(\beta)&=&e^{-i\phi_2F_{z}^{jk}}e^{-i\beta\,F_{x}^{jk}}e^{i\phi_2F_{z}^{jk}}\,.
\end{eqnarray}

All experiments in this contribution were performed at a
temperature of $T=50$~K.

    \begin{figure}[htb]
\includegraphics[width=0.4\textwidth]{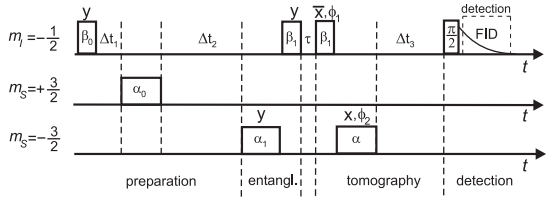}
        \caption{Pulse sequence for preparation and detection of the pseudo entangled state
        $\Psi^{-}$ in $\mathrm{^{15}N@C_{60}}$. The left part refers to the preparation of the pseudo pure state.
        The middle part represents the preparation and detection of the entangled state.
The FID after the selective ESR pulse (right hand side) serves as a monitor signal.
The angles $\varphi_1$ and $\varphi_2$ of the tomography sequence are varied in increments (see text)
and are used for separating signals of entangled states from those of individual spins.
The waiting times between the pulses were chosen $\Delta t_1=5\,\mu$s, $\Delta t_2=100\,\mu$s,
$\Delta t_3=150\,\mu$s and $\tau=40$~ns (see text). }
        \label{entpulse}
    \end{figure}

These pulses serve different purposes as is sketched in the general pulse sequence displayed in Fig. \ref{entpulse}.
The total sequence is separated into four segments, namely (1) preparation of pseudo pure states, (2) creation of entanglement,
(3) state tomography and (4) detection via the electron spin free induction decay (FID).

\section{Pseudo pure states.}\label{sec:pseudopure}

\subsection{Preparation of pseudo pure states}
    Like in liquid state NMR quantum computing \cite{cory:97,gershenfeld:97a,knill:98a} we are dealing here with mixed quantum states and moreover we start initially from a thermal equilibrium state, namely the Boltzmann state which can be expressed as
    \begin{equation}\label{eqn:rhoB}
        \rho_{\mathrm{B}}=
        \frac{\mathrm{e}^{-\beta_\mathrm{B} \omega_S S_z}}
        {\mathrm{Tr}\left\{\mathrm{e}^{-\beta_\mathrm{B} \omega_S S_z}\right\}}
    \end{equation}

with $\beta_\mathrm{B} = \hbar/k_\mathrm{B}T$ and where we have
applied the approximation $\omega_S \gg \omega_I, a$. With all
experiments performed at $T=50$~K and
$\hbar\omega_S/(k_{\mathrm{B}}T)\approx0.01\ll1$ the expression
can be further simplified by applying the high temperature
approximation which results for our $S=3/2$ and $I=1/2$ system in

    \begin{equation}\label{eqn:rhoB}
        \rho_{\mathrm{B}}\approx\frac{1}{8}\left(I_8-K_BS_z\right)
        \quad\mbox{with}\quad
        K_B=\frac{\hbar\omega_S}{k_{\mathrm{B}}T}\, .
    \end{equation}
    and where $I_8$ represents the $8 \times 8$ identity matrix. Eq.(\ref{eqn:rhoB}) can be
    rewritten as
     \begin{eqnarray}\label{eqn:rhoBP}
    \rho_{\mathrm{B}}&\approx&
    \frac{1}{8}\left(1-\frac{1}{2}K_B\right)I_8+\frac{1}{4}K_B\rho_{\mathrm{P}}\quad\mbox{with}\quad\\
    \rho_{\mathrm{P}}&=&\frac{1}{4}I_8-\frac{1}{2}S_z.
    \end{eqnarray}

    This defines the $8\times 8$ so-called pseudo Boltzmann matrix
$\rho_{\mathrm{P}}$ which will be the starting point for
preparing pseudo pure initial density matrices.

We treat in the following the preparation of the pseudo pure
density matrix $\rho_{\mathrm{P}10}$, from which we will extract
finally the two qubit density matrix $\rho_{10}$ with diagonal
components $\{0,0,1,0\}$, as an example. It is important to
prepare a pseudo pure state which not only resembles the
corresponding pure quantum state with the same operator structure
but also corresponds to a large nuclear spin alignment which is a
prerequisite for reaching a high degree of entanglement.
According to Fig. \ref{entpulse} we first apply a pulse
$P_{y}^{(-)}(\beta_0)$ with
$\beta_0=\arccos(-1/3)=109,47^{\circ}$ at the electron spin
transitions for $m_I=-1/2$. After a waiting time of
$\tau_1=5\,\mu$s all transverse components have decayed and a
pulse $P_{y}^{(12)}(\alpha_0)$ pulse with $\alpha_0=\pi/2$ at the
$1\leftrightarrow 2$ ENDOR transition follows in order to
equalize the populations of level 1 and 2. After an additional
waiting time of $\tau_2=100\,\mu$s all transient components of
the density matrix have decayed and the pseudo pure state
$\rho_{\mathrm{P}10}$ has been reached with

\begin{equation}\label{eqn:diagrhoP10}
    \rho_{\mathrm{P}10} = \left\{\mathbf{0},\mathbf{0},~0,\frac{1}{3},~\frac{1}{2},\frac{1}{6},\mathbf{1},\mathbf{0}\right\}
\end{equation}

 where here and in the following we express diagonal density matrices by the diagonal element vector
 where the elements in bold type correspond to the matrix $\rho_{10}$ of
 the subsystem $m_S=\pm 3/2$ and $m_I=\pm 1/2$.

The pseudopure state $\rho_{11}$ can be prepared in a similar way
by exciting the electron spin transition with $m_I=+1/2$.

    \begin{figure}[htb]
        \centerline{\includegraphics[width=0.3\textwidth]{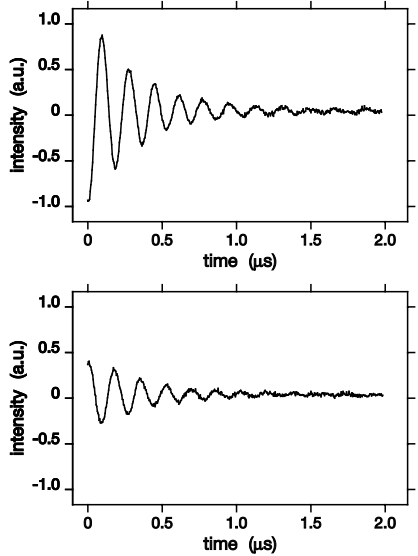}}
        \caption{Rabi oscillations at the electron spin transitions with $m_I=-1/2$.
         Top: Boltzmann state $\rho_{\mathrm{P}}$. Bottom: After application of a $P_y^{(-)}(\beta_0)$ pulse.}
        \label{tomorho10esr}
    \end{figure}

\subsection{Tomography of pseudo pure states}

The reconstruction of the density matrix of a quantum state by
density matrix tomography has already been applied in liquid
state NMR quantum computing
\cite{chuang:98b,knill:98a,laflamme:98,boulant:02,teklemariam:02}).
Here the situation is different because we are dealing with ESR
and ENDOR transitions in a solid at low temperatures. In order to
verify the proper preparation of the pseudo pure density matrices
we have applied Rabi oscillation measurements on different
electron and nuclear transitions. This is shown for the
pseudopure state $\rho_{10}$ in Fig. \ref{tomorho10esr} as an
example. There the magnitude of the electron spin free induction
decay (FID) signal is plotted after applying pulses of variable
length at particular ESR or ENDOR transitions. The Rabi
oscillations decay due to the microwave- or rf-field
inhomogeneity. However, their amplitude and initial phase
represent the population difference in magnitude and sign of the
particular transition where the pulse is applied. The amplitude
of the Rabi oscillations was determined by Fourier transformation
and integration over the line in the corresponding spectrum. The
experimental values were calibrated with respect to the Boltzmann
state $\rho_{\mathrm{P}}$ whose amplitude is defined as
$A_{\mathrm{P}}\equiv 1/2$. Similar measurements were obtained
for state $\rho_{11}$.

Figure \ref{tomorho10esr} shows Rabi oscillations at the ESR
transition with $m_I=-1/2$. A comparison is made between the
Boltzmann state (top) and an inverted state (bottom) after the
application of a $P_y^{(-)}(\beta_0)$ pulse. Note the change in
amplitude and sign with respect to the Boltzmann state. The level
of inversion was determined to -0.329 which results in an angle $
\beta_0=\arccos(-0.329)\triangleq 109.2^{\circ}$ to be compared
with the ideal value $\beta_0=\arccos(-1/3)\triangleq
109.5^{\circ}$.

In addition we performed similar Rabi precession experiments at
the ENDOR transitions $1\leftrightarrow2$ and $7\leftrightarrow8$
of the pseudo pure state $\rho_{\mathrm{P}10}$. From the
amplitude of the Rabi oscillations we have determined the
corresponding population differences. The deviation from the
theoretically expected values can be expressed in terms of
deviations of the angles $\beta_0$ and $\alpha_0$ from their
ideal values. By comparing the expected signal strength

    \begin{equation}
        A_{\mathrm{P}10}^{(12)}=-\frac{9}{80}\cos\alpha_0\left(1-\cos\beta_0\right)\,
    \end{equation}

with and without the $P_y^{(12)}(\alpha_0)$ pulse leads to an
effective value of $ \alpha_0=1.55\equiv 88.8^{\circ}$ which
reflects a rather small deviation from the ideal value
$\alpha_0=\pi/2\equiv 90.0^{\circ}$. Deviations of the angles
$\alpha_0$ and $\beta_0$ from their ideal values are not due to
misadjusting but are rather caused by microwave- and rf-field
inhomogeneities. This affects the diagonal elements of
$\rho^{(exp)}_{P10}$ and leads to the following values

\begin{equation}\label{eqn:diagrhoP10exp}
   \rho^{(exp)}_{\mathrm{P}10} = \left\{\mathbf{-0.01},\mathbf{0.01},0,~0.33,~\frac{1}{2},~0.17,\mathbf{1},\mathbf{0.00}\right\}
\end{equation}

where negative values arise from experimental errors. After such
a data analysis we obtain for the pseudo pure fictitious two
qubit density matrix $\rho_{10}$
    \begin{equation}
    \rho_{10}=
        \left(
        \begin{array}{cccc}
        -0.01&0&0&0\\
        0&0.01&0&0\\
        0&0&1&0\\
        0&0&0&0.00
        \end{array}
        \right)\,.\label{eqn:rho10}
    \end{equation}
The errors are in the last digit. Note that the numbers 1 and
$\frac{1}{2}$ derive from $\rho_{\mathrm{P}}$ of the Boltzmann
state (see Eq.(\ref{eqn:rhoBP})) and are not changed by the
preparation pulses. Matrix elements of those types are written in
the following as bare numbers or fractions. In a similar way as
the pseudo pure density matrix $\rho_{11}$ was obtained.

These pseudo pure states serve as initial states for the
generation of entangled states. $\rho_{10}$ was used to generate
the Bell-states
    $\Psi_{27}^{\pm}$ and $\rho_{11}$ to generate $\Phi_{18}^{\pm}$.

\section{Pseudo entangled states}
\subsection{Preparation of pseudo entangled states}

    Pseudo entangled states were prepared according to the pulse sequence depicted in Fig. \ref{entpulse} (central part). Instead of the usual Hadamard transformation a selective $\pi/2$ pulse on a specific nuclear spin transition was applied which has a related effect as the Hadamard transform. The CNOT-gate was implemented by a
    selective $\pi$-pulse on an electron spin transition. Either $\Psi^{\pm}$ or $\Phi^{\pm}$ states were generated depending on the pseudo pure initial state and the transitions used. In order to distinguish the different Bell states we have used phase rotations of the pulses in the tomography sequence (see Fig. \ref{entpulse}) as presented before \cite{mehring:03}.

    According to equations (\ref{psi27}) and (\ref{phi18}) we need to prepare the following density matrices
    \begin{equation}
        \rho_{\Psi^{\pm}}^{27}=\frac{1}{2}\left|2\pm7\right\rangle\left\langle2\pm7\right|
        \quad \mbox{and}\quad
        \rho_{\Phi^{\pm}}^{18}=\frac{1}{2}\left|1\pm8\right\rangle\left\langle1\pm8\right|\,.\label{psipm27}
    \end{equation}

    Starting from the pseudo pure state $\rho_{\mathrm{P}10}$ first a $\alpha_1=\pi/2$-pulse was applied at the nuclear spin  transition $7\leftrightarrow8$ followed immediately by a $\beta_1= \pi$-pulse at electron spin
    transitions with $m_I=-1/2$ (see Fig. \ref{entpulse}) leading to the following sequence of unitary transformations
    \begin{equation}
        \rho_{\mathrm{P}\Psi^{\pm}}^{27}=U_{\mathrm{\pm}}^{27}\rho_{\mathrm{P}10}\left(U_{\mathrm{
        \pm}}^{27}\right)^\dagger \quad\mbox{with}\quad U_{\mathrm{\pm}}^{27}=
        P_{y}^{-}(\mp\pi)\,P_{y}^{78}(+\pi/2)\, .\label{psiprep}
    \end{equation}

Under ideal conditions this would result in
    \begin{equation}
    \rho_{\mathrm{P}\Psi^{\pm}}^{27}=\left(
        \begin{array}{cccccccc}
        \mathbf{0}&\mathbf{0}&0&0&0&0&\mathbf{0}&\mathbf{0}\\
        \mathbf{0}&\mathbf{\frac{1}{2}}&0&0&0&0&\mathbf{\pm\frac{1}{2}}&\mathbf{0}\\
        0&0&0&0&0&0&0&0\\
        0&0&0&\frac{1}{6}&0&0&0&0\\
        0&0&0&0&\frac{1}{2}&0&0&0\\
        0&0&0&0&0&\frac{1}{3}&0&0\\
        \mathbf{0}&\mathbf{\pm\frac{1}{2}}&0&0&0&0&\mathbf{\frac{1}{2}}&\mathbf{0}\\
        \mathbf{0}&\mathbf{0}&0&0&0&0&\mathbf{0}&\mathbf{0}\\
        \end{array}
        \right)\,
    \end{equation}

    where the elements in bold type again belong to the fictitious two qubit
    submatrix

    \begin{equation}
    \rho_{\Psi^{\pm}}=\left(
        \begin{array}{cccc}
        0&0&0&0\\
        0&\frac{1}{2}&\pm\frac{1}{2}&0\\
        0&\pm\frac{1}{2}&\frac{1}{2}&0\\
        0&0&0&0\\
        \end{array}
        \right)\,. \label{psipm}
    \end{equation}

   The two qubit submatrix of the density matrix $\rho_{\mathrm{P}\Psi^{\pm}}^{27}$
    is almost identical to $\rho_{\Psi^{\pm}}^{27}$ (see eq.
    (\ref{psipm27})) except for additional diagonal elements
    outside the fictitious two qubit subsystem.

    Similarly we have prepared the $\rho_{\Phi^{\pm}}^{18}$ states by starting from the pseudo pure state $\rho_{\mathrm{P}11}$ and applied a $\pi/2$-pulse on nuclear transition $7\leftrightarrow 8$ followed immediately by a $\pi$-Pulse on electron spin transitions with $m_I=+1/2$
    \begin{equation}
        \rho_{\mathrm{P}\Phi^{\pm}}^{18}=U_{\mathrm{\pm}}^{18}\rho_{\mathrm{P}11}
        \left(U_{\mathrm{\pm}}^{18}\right)^\dagger \quad\mbox{with}\quad
        U_{\mathrm{\pm}}^{18}=P_{y}^{+}(\pm\pi)\,P_{y}^{78}(+\pi/2)\, \label{phiprep}
    \end{equation} with
    \begin{equation}
    \rho_{\mathrm{P}\Phi^{\pm}}^{18}=\left(
        \begin{array}{cccccccc}
        \mathbf{\frac{1}{2}}&\mathbf{0}&0&0&0&0&\mathbf{0}&\mathbf{\pm\frac{1}{2}}\\
        \mathbf{0}&\mathbf{0}&0&0&0&0&\mathbf{0}&\mathbf{0}\\
        0&0&\frac{1}{6}&0&0&0&0&0\\
        0&0&0&0&0&0&0&0\\
        0&0&0&0&\frac{1}{3}&0&0&0\\
        0&0&0&0&0&\frac{1}{2}&0&0\\
        \mathbf{0}&\mathbf{0}&0&0&0&0&\mathbf{0}&\mathbf{0}\\
        \mathbf{\pm\frac{1}{2}}&\mathbf{0}&0&0&0&0&\mathbf{0}&\mathbf{\frac{1}{2}}\\
        \end{array}\label{rhopsipm}
        \right)\,
    \end{equation}
    where the elements in bold type again belong to the fictitious two qubit
    submatrix
    \begin{equation}
    \rho_{\Phi^{\pm}}=\left(
        \begin{array}{cccc}
        \frac{1}{2}&0&0&\pm\frac{1}{2}\\
        0&0&0&0\\
        0&0&0&0\\
        \pm\frac{1}{2}&0&0&\frac{1}{2}\\
        \end{array}
        \right)\,.\label{phipm}
    \end{equation}

    Here we have used a pulse length of $1.6\, \mu$s for the $\pi/2$ radio frequency pulse and 88~ns for
    the microwave $\pi$-pulse. By this entanglement was prepared within approximately $1.7\,\mu$s starting
    from the pseudo pure state.

    Note that the density matrices $\rho_{\Psi^{\pm}}$ (Eq. (\ref{psipm})) and
    $\rho_{\Phi^{\pm}}$ (Eq. (\ref{phipm})) correspond to the Bell states defined in Eq. (\ref{bell}). Here we investigated which density matrices are obtained under ideal conditions. In the next section we analyze the experimentally obtained density matrices by performing a density matrix tomography.

\subsection{Density matrix tomography}
In the density matrix tomography used here we combine Rabi precession to determine the populations of the quantum levels together with phase incrementation measurements to discriminate between different off-diagonal components.

Since entangled states are not directly observable we need to transform them to an observable state.  We therefore apply an {\em entangled state detector} consisting for the states $\Psi^{\pm}$ the pulse sequence $P^{(-)}_x(-\beta_1)$ with $\beta_1=\pi$ followed by $P^{(78)}_x(\alpha)$ with $\alpha=\pi/2$ (see Fig. \ref{entpulse}). We note that we always use the same angle $\beta_1$ in the preparation and the detection sequence or technically speaking we use the same amplitude and pulsewidth.

\subsubsection{Phase rotation of entangled states}
An essential aspect of our type of tomography for distinguishing between the different Bell states is their dependence on $z$-rotations as discussed in section \ref{z-rot} \cite{mehring:03}. Accordingly we have implemented the detection sequence with $P^{-}_x(-\pi,~\phi_1)$ and $P^{78}_x(\pi/2,~\phi_2)$ pulses.
 The corresponding unitary transformation of the detection sequence then reads
    \begin{equation}
    U^{27}_{\mathrm{d}}(\phi_1,~\phi_2)= P^{78}_x(\pi/2,~\phi_2)\,P^{-}_x(-\pi,~\phi_1).
    \end{equation}

The measured quantity after the detection is the $z$-magnetization of the electron spin subsystem with $m_I = -1/2$ in the case of $\Psi_{27}^{\pm}$. The observed signal strength is therefore expected to vary as
    \begin{eqnarray}
    S_{\Psi}^{\pm}(\phi_1,\phi_2) &=& -\frac{\mathrm{Tr}\left\{F^{-}_z\,U^{27}_{\mathrm{d}}\,
\rho_{\mathrm{P}\Psi^{\pm}}^{27}\,U^{27\dag}_\mathrm{d}\right\}}{\mathrm{Tr}\left\{\left(F^{-}_z\right)^2\right\}}\\
    &=& \frac{2}{15}\pm\frac{3}{20}\cos(3\phi_1-\phi_2)\label{psidet}
    \end{eqnarray}
    where $F^{-}_z$ is the fictitious electron spin 3/2 of the $m_I=-1/2$ subsystem. The phase dependence with
    $(3\phi_1-\phi_2)$ is characteristic for the $\Psi^{\pm}$ states. The variation with $3\phi_1$ originates from   the $\pm 3/2$ levels.

    Similarly we have used the sequence
    \begin{equation}
    U^{18}_{\mathrm{d}}(\phi_1,~\phi_2)= P^{78}_x(\pi/2,~\phi_2)\,P^{+}_x(-\pi,~\phi_1).
    \end{equation}
    for the detection of the $\Phi_{18}^{\pm}$ states resulting in a detector signal
    \begin{eqnarray}
    S_{\Phi}^{\pm}(\phi_1,\phi_2) &=& -\frac{\mathrm{Tr}\left\{F^{+}_z\,U^{18}_{\mathrm{d}}\,    \rho_{\mathrm{P}\Phi^{\pm}}^{18}\,U^{18\dag}_\mathrm{d}\right\}}{\mathrm{Tr}\left\{\left(F^{+}_z\right)^2\right\}}\\
    &=& \frac{2}{15}\pm\frac{3}{20}\cos(3\phi_1+\phi_2)\label{phidet}
    \end{eqnarray}
    for $\Phi_{18}^{\pm}$ and where preparation and detection pulses were applied at the $m_I=+1/2$ ESR transitions.  The phase variation with $(3\phi_1+\phi_2)$ is characteristic for the $\Phi_{18}^{\pm}$ states.

\begin{figure}[htb]        \centerline{\includegraphics[width=0.2\textwidth]{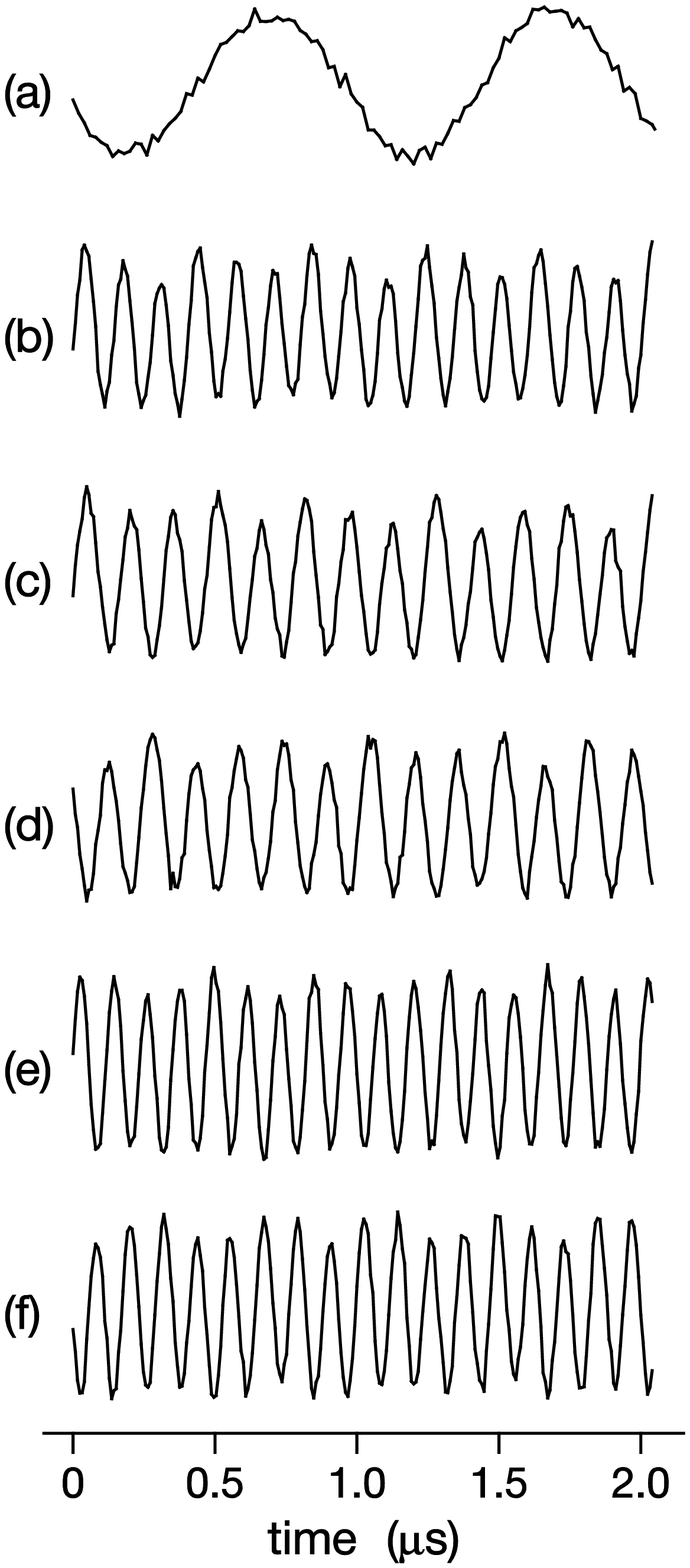}
        \quad\includegraphics[width=0.2\textwidth]{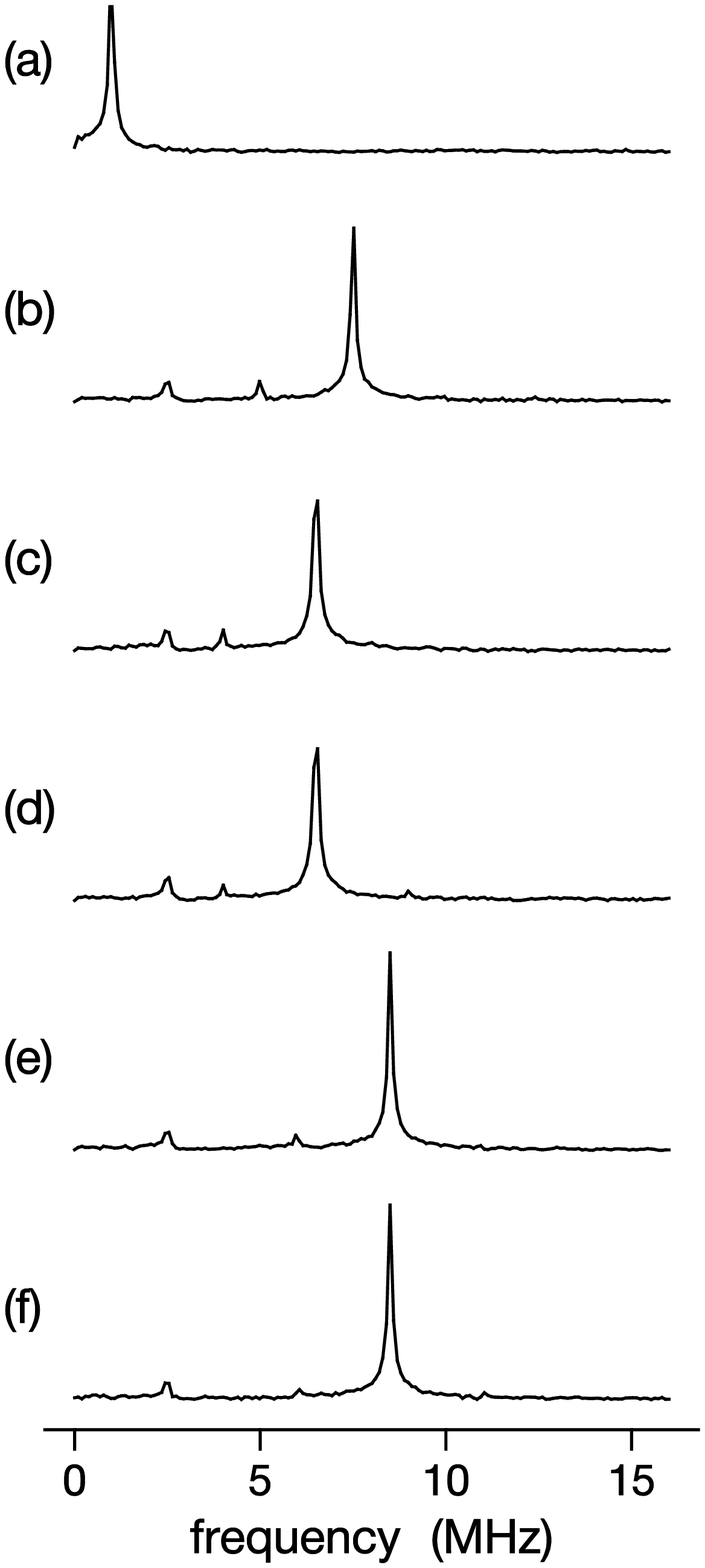}}
        \caption{Left: Phase interferograms for a complete set of Bell states according to the pulse sequence in
         Fig.\ref{entpulse}  with phase incrementation frequencies $\nu_1=2.5$~MHz and
        $\nu_1=1.0$~MHz. The experimental traces belong to $\rho_{\mathrm{P}\Psi^{-}}^{27}$ (a) to (c),
         $\rho_{\mathrm{P}\Psi^{+}}^{27}$(d), $\rho_{\mathrm{P}\Phi^{+}}^{18}$ (e) and
          $\rho_{\mathrm{P}\Phi^{-}}^{18}$ (f). In the first two interferograms only one of
        phase angle was incremented:
        (a): incrementation of $\phi_2$, (b): incrementation of
        $\phi_1$. In the interferograms (c) to (f) both phase
        angles $\phi_1$ and $\phi_2$ were incremented simultaneously.
        Right: Corresponding power spectra after Fourier transformation .
        There is a dominant line at $3\nu_1-\nu_2$ for the $\rho_{\mathrm{P}\Psi^{\pm}}^{27}$ states and
        at $3\nu_1+\nu_2$ for the $\rho_{\mathrm{P}\Phi^{\pm}}^{18}$ states. See text for additional weak lines.
       }\label{psimpphipm}
    \end{figure}

    The phase dependence of the entangled state was measured by repeating the whole pulse sequence shown in Fig. \ref{entpulse} for different values of $\varphi_1$ and $\varphi_2$. The angles $\varphi_1$ and $\varphi_2$ were incremented in steps of $\Delta\varphi_1$ and $\Delta\varphi_2$, leading to phase angles
    \begin{eqnarray}
        \varphi_1(n)&=&n\, \Delta\varphi_1 =n\nu_1\, \Delta t\\
        \varphi_2(n)&=&n\, \Delta\varphi_2 = n\nu_2\, \Delta t
    \end{eqnarray}
    after $n$ steps. We have introduced here a virtual time scale $t=n\, \Delta t$
   which defines the virtual frequencies $\nu_1$ and $\nu_2$. By incrementing both phase angles $\phi_1$ and $\phi_2$ simultaneously an oscillatory behavior of the detection signal according to eqns.
    (\ref{psidet}) and (\ref{phidet}) is expected. In the following we call the corresponding traces, shown in Fig.(\ref{psimpphipm}) phase interferograms.

    Figure (\ref{psimpphipm}) shows phase interferograms and the
    corresponding Fourier spectra for phase incrementation
   the two frequencies $\nu_1=2.5$~MHz and $\nu_2=1.0$~MHz. The presence of entangled
    states is revealed by dominant lines at $3\nu_1-\nu_2$ for the
    $\rho_{\mathrm{P}\Psi^{\pm}}^{27}$ states ((c) and (d)) and at $3\nu_1+\nu_2$ for the
    $\rho_{\mathrm{P}\Phi^{\pm}}^{18}$ states ((e) and (f)) in
    contrast to the single spin phase variations seen in (a) and (b).

    The additional weak lines appearing at $\nu_1$ and $2\nu_1\mp\nu_1$ are artifacts due to pulse imperfections.
    A theoretical analysis shows that the ratio of the amplitudes of the lines at
    $2\nu_1\mp\nu_1$ and $3\nu_1\mp\nu_1$ depends on the deviation $\epsilon_{\beta}$ of the electron spin rotation angle $\beta_1$ from $\pi$

    \begin{equation}
\frac{A(2\nu_1\mp\nu_2)}{A(3\nu_1\mp\nu_2)}\approx24\frac{\epsilon_{\beta}^2}{\left(16-21\epsilon_{\beta}^2\right)}
            \quad\mbox{with}\quad \epsilon_{\beta}=\beta_1-\pi\,.
    \end{equation}
    where terms up to second order in $\epsilon_{\beta}$ were
    considered. This we have used to quantify the precision of the
    applied electron spin pulse $\beta_1$.
    For example the mean deviation from the ideal situation for state $\rho_{\mathrm{P}\Psi^-}^{27}$ (Fig. \ref{psimpphipm} (c)) can be determined to be $\left|\overline{\epsilon}_{\beta}\right|=0.23$.
    This is in accordance with the fact that a perfect $\pi$-pulse does not exist, because microwave field
    inhomogeneities alone already prevent a complete inversion. $\overline{\epsilon}_{\beta}$ therefore comprises
    positive as well as negative deviations. Smaller values of $\overline{\epsilon}_{\beta}$ could therefore be achieved by more homogeneous microwave fields. The determination of $\overline{\epsilon}_{\beta}$ as a measure of the precision of the entangled states was performed also for the other Bell states and is an essential part for the estimation of the experimental precision of the created states.

\subsubsection{The diagonal part}
The diagonal part of the density matrix of the entangled states was determined by Rabi oscillations in a similar way as already discussed for the pseudo pure states in section \ref{sec:pseudopure}. Here we combine electron spin Rabi oscillations together with nuclear spin Rabi oscillations at the relevant transitions $1\leftrightarrow 2$ and $7\leftrightarrow 8$.

    After the preparation of the entangled state a waiting time of $100\,\mu$s was added to let all off-diagonal elements decay. After this a pulse of variable length at the particular transition was added resulting in the observed Rabi oscillations where the initial amplitude is proportional to the difference of the corresponding diagonal matrix elements. The results were calibrated with respect to the previous data obtained for the Boltzmann and pseudo pure state.
    By applying the appropriate numerical analysis we extracted for the $\rho_{\mathrm{P}\Psi^-}^{27}$ state the following values for the effective rotation angles $\alpha_1^{-}=86.6^{\circ}$, $\alpha_1^{12}=86.8^{\circ}$  and $\alpha_1^{78}=88.7^{\circ}$. In this analysis we included the mean deviation $|\overline{\epsilon}_{\beta}| = 0.23$  which is legitimate since $\epsilon_{\beta}$ enters the equations only in even powers. Since all three measurements are equally sensitive to experimental
    errors we have taken their mean value
    \begin{equation}
        \alpha_1=\frac{1}{3}\left(\alpha_1^{-}+\alpha_1^{12}+\alpha_1^{78}\right)=87.4^{\circ}\,.
    \end{equation}

    By combining the previous results for the initial pseudo pure state
    $\rho_{10}$ from Eq.(\ref{eqn:diagrho}) with the values of $\alpha_1$ and
    $|\epsilon_{\beta}|$ we obtain the complete set of the diagonal elements $r_j^\Psi$ of the density
    matrix $\rho_{P\Psi^-}^{27}$ as
    \begin{equation}
    r_j^\Psi = \left\{\mathbf{-0.01},~\mathbf{0.47},~0,~0.19,~\frac{1}{2},~0.31,~\mathbf{0.52},~\mathbf{0.02}\right\}.
    \end{equation}

    The diagonal elements of the corresponding sublevel density
    matrix $\rho_{\Psi^-}$ are printed in bold type. Their experimental error are estimated to be equal or smaller
    than $\pm0.05$. The values are therefore in good agreement with the ideal density matrix in
    Eq. (\ref{rhopsipm}). Elements originating from the initial Boltzmann density matrix are unaffected by the
    pulse sequence are presented by bare numbers or fractions. A similar analysis was made for the diagonal elements
    of the other Bell states. We remark that this analysis is affected by the decay time of the diagonal terms during the waiting time of $100\,\mu$s introduced between preparation and detection. In order to estimate its influence we measured this decay time to $2.60$~ms which results in a decay by only 4\%. We have not corrected
for this decay but considered it to be within the experimental error.

\subsubsection{The off-diagonal part}\label{sec:offdiag}
    Because we have experimentally determined already both pulse angles of the sequence for preparing the entangled states, namely $\alpha_1$ and $\beta_1=\pi+\epsilon_{\beta}$, the unitary transformations for the preparation of entanglement are known and could in principle be applied to calculate the corresponding off-diagonal elements of the density matrix. In order to experimentally determine the off-diagonal elements we have, however, applied
an alternative tomography sequence to determine the actual values of the off-diagonal elements with respect to the diagonal elements.

\begin{figure}[htb]
        \centerline{\includegraphics[width=0.4\textwidth]{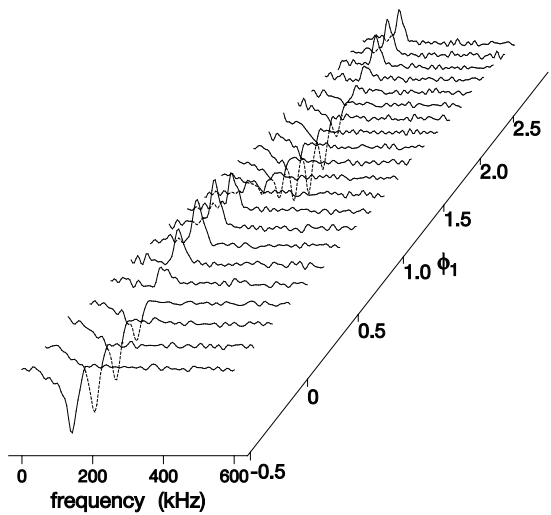}}
        \caption{A series of Fourier transformed Rabi oscillations at the $7\leftrightarrow8$ transition of the
        $\rho_{\mathrm{P}\Psi^-}^{27}$ state for different values of $\phi_1$.}\label{racopsim}
    \end{figure}

In the tomography part of the sequence in Fig. \ref{entpulse} we have incremented the rotation angle $\alpha$ (Rabi oscillation) at the $7\leftrightarrow8$ transition for different settings of the phase $\phi_1$ and a constant value $\phi_2$. After Fourier transform of the Rabi oscillations spectra result where their real parts are displayed in Fig. \ref{racopsim} for different values of $\phi_1$. Note, however, that the signal now depends on diagonal as well as on off-diagonal elements of the density matrix.

Except for an additional $\phi_2=90^{\circ}$ phase rotation, in order to be sensitive to "$\sin\alpha$"-terms, the amplitudes were evaluated in the same way as discussed for the Rabi oscillations.

    \begin{figure}[htb]
        \centerline{\includegraphics[width=0.35\textwidth]{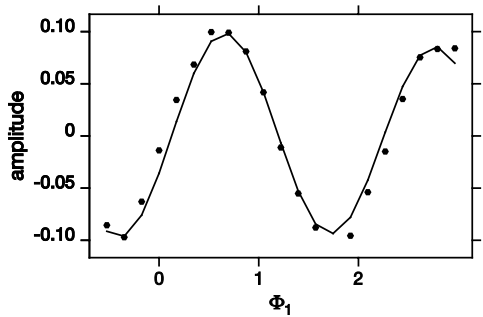}}
        \caption{Amplitudes variation of the Rabi spectra of  Fig. \ref{racopsim} versus $\phi_1$.}\label{rapsimfit}
    \end{figure}

A simple calculation leads to the following expression
     \begin{multline}\label{eqn:sinalphatomo}
        S_{\Psi^-}^{\alpha}=B_0+A_1 \cos{\alpha}-A_2 \sin{\phi_1}+\\A_3\sin\alpha
        \sin(2\phi_1-\phi_2)-A_4 \sin\alpha
        \cos(3\phi_1-\phi_2)\,.
    \end{multline}

for the detector signal of this type of tomography. Due to the
appropriate phase setting combined with selecting the relevant
terms of corresponding spectrum after a complex Fourier transform
allows us extract only terms with "$\sin\alpha$" i.e. those two
terms with amplitudes $A_3$ and $A_4$ in Eq.
(\ref{eqn:sinalphatomo}) in the data shown in Fig.
\ref{rapsimfit}. By using  the ratio $A_3/A_4$ from the phase
interferogram in
    Fig. \ref{psimpphipm} (c) we were able to determine the amplitude $A_4$ from a fit to the data shown in Fig. \ref{rapsimfit}.

The most important matrix element of the entangled state $\rho_{\mathrm{P}\Psi^-}^{27}$, namely
the matrix element $r_{27}=r_{72}$ enters the $A_4$ parameter in the following way
    \begin{equation}
    A_4=-\frac{3}{10}r_{27}\left(\cos\left(\frac{\epsilon_{\beta}}{2}\right)\right)^3
    \,\,\mbox{with}\,\,\epsilon_{\beta}=\overline{\epsilon}_{\beta}=\pm0.23.
    \end{equation}
    From the knowledge off all other parameters we obtain the value.
    \begin{equation}\label{r27}
        r_{27}=r_{72}=-0.31\pm{0.04}\,.
    \end{equation}

    This is only $0.64\%$ of the theoretically expected value since from the experimentally known values of the angles $\beta_1$ and $\alpha_1$ of the preparation sequence and the pseudo pure density matrix one would expect
$r_{27}^{\mathrm{th}}=r_{72}^{\mathrm{th}}=-0.49$. The reason for the reduced value is the decoherence of the
    entangled state during and after preparation and the delay (including finite pulse width) up to the tomography.

\subsubsection{Decoherence}
Decoherence reduces the off-diagonal values of the density matrix since the pulse sequence for the preparation and tomography requires finite pulse lengths and a minimum delay $\tau$ between the excitation and the tomography sequence.

    \begin{figure}[htb]
        \centerline{\includegraphics[width=0.35\textwidth]{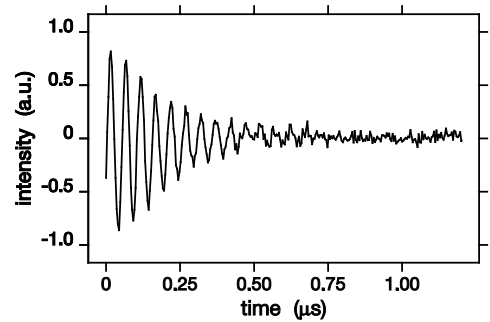}}
        \caption{Decay of the off-diagonal element $r_{27}$ for state $\rho_{P\Psi^-}^{27}$ modulated by
        double resonance TPPI (time proportional phase incrementation) with frequencies $\nu_1=8.0$~MHz,
         $\nu_2=4.0$~MHz. The decay is modulated by the characteristic frequency $3\nu_1-\nu_2$.}\label{en069s}
    \end{figure}

We have measured this decay for the $\Psi^-$ state by incrementing the delay time between the preparation sequence and the tomography sequence. The decoherence decay shown in Fig. \ref{en069s} is modulated by the characteristic frequency $3\nu_1-\nu_2$ for the $\rho_{\mathrm{P}\Psi^-}^{27}$ state imposed by the time proportional phase incrementation (TPPI) procedure \cite{drobny:79, bodenhausen:80}. Our type of double resonance TPPI was implemented by combined phase increments of $\phi_1$ and $\phi_2$ in the tomography sequence. In order to purify the decay function from artifacts corresponding to the weak lines in Fig. \ref{psimpphipm} (c) we have added measurements with $y$ and $-y$ preparation pulses (phase cycling). As a consequence the decay function is solely modulated by the frequency $3\nu_1-\nu_2$ characteristic for the entangled state $\Psi^-$. Other coherent excitations would appear with
other phase incrementation frequencies.

By analyzing the decay function in Fig. \ref{en069s} we were able
    to determine the time constant of the decoherence of the entangled state to
    \begin{equation}
        T_{\Psi}=(208\pm10)\,\mbox{ns.}
    \end{equation}

Since the process of decoherence is not the subject of this contribution we only remark that it is dominated by an inhomogeneous distribution of the ESR transitions. In order to reconstruct the values of $r_{27} = r_{72}$ right after their preparation we need to evaluate the effective delay time of the tomography sequence. It is comprised of finite pulse width as well as actual delay $\tau$ due to technical constraints. Since the ESR pulses had to be soft in order to avoid cross talk with the other ESR line the corresponding pulse width was $88$~ns which is not short with respect to $T_{\Psi}$. By considering both, the finite pulse width and the delay time $\tau$ we could reconstruct the initial values of $r_{27} = r_{72}$ to

    \begin{equation}
     \left.r_{27}^{\Psi_-}\right|_{\mathrm{initial}}=\left.r_{72}^{\Psi_-}\right|_{\mathrm{initial}}=-0.42
    \end{equation}

This gives a reasonably high degree of entanglement right after creation.

\subsubsection{The complete density matrix}
\begin{figure}[htb]
        \centerline{\includegraphics[width=0.3\textwidth]{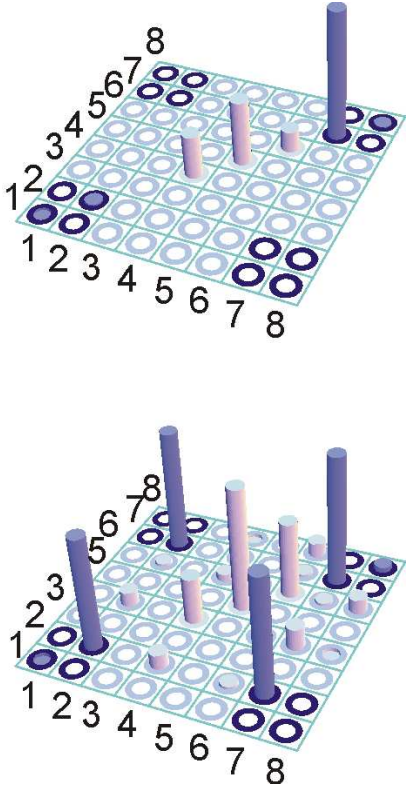}}
        \caption{Top: Pseudo pure initial density matrix $\rho_{P10}$. Bottom: Density matrix $\rho_{\mathrm{P}\Psi^{+}}^{27}$ }\label{fig:rhoP10Psip}
    \end{figure}

The summary of the density matrix tomography presented here
results in the complete density matrix for all four Bell states.
Fig. \ref{fig:rhoP10Psip} shows a graphical representation of the
complete density matrix of $\rho_{\mathrm{P}\Psi^{+}}^{27}$
together with complete initial density matrix $\rho_{P10}$.

The elements of the two qubit submatrix are enhanced by a darker representation, whereas the other matrix elements are presented lighter. Note that the initial matrix $\rho_{P10}$ already contains diagonal elements in the middle area which are not relevant for the subsystem we are interested in. They are only slightly affected by the entanglement sequence. The errors of the diagonal elements of $\rho_{\mathrm{P}\Psi^{+}}^{27}$ were estimated to be smaller or equal to $\pm0.05$. The errors of the nominally zero off-diagonal elements are typically smaller than $\pm0.06$. The nominally zero off-diagonal elements in the two qubit submatrix have errors on the order of $\pm0.01$. Some of the density matrix elements are unaffected by the preparation sequence are identically zero.

Similar results were obtained for $\rho_{\mathrm{P}\Psi^{+}}^{27}$ and with $\rho_{P11}$ as initial density matrix for the $\rho_{\mathrm{P}\Phi^{\pm}}^{19}$ Bell states. This leads us finally to the following two qubit submatrices corresponding to the two qubit Bell states.

            \begin{equation}
            \rho_{\Psi^-}=\left(
            \begin{array}{cccc}
                -0.01&0&0&0\\
                0&+0.47&-0.42&0.00\\
                0&-0.42&+0.52&0.00\\
                0&0.00&0.00&0.02
                \end{array}
                \right)\,.
            \end{equation}

Off-diagonal elements at detection time: $r_{27}^{\Psi^-}=r_{72}^{\Psi^-}=-0.31\pm0.04$.
Quoted values were reconstructed at preparation time due to the measured decoherence time $T_{\Psi^-}=208\pm10\,\mbox{ns.}$

    \begin{equation}
        \rho_{\Psi^+}=\left(
        \begin{array}{cccc}
        -0.01&0&0&0\\
        0&+0.46&+0.42&0.00\\
        0&+0.42&+0.53&0.00\\
        0&0.00&0.00&+0.02
        \end{array}\right)\,.
    \end{equation}

Off-diagonal elements at detection time: $r_{27}^{\Psi^+}=r_{72}^{\Psi^+}= 0.31\pm0.04$.
Quoted values were reconstructed at preparation time due to the measured decoherence time $T_{\Psi^+}=198\pm10\,\mbox{ns.}$

        \begin{equation}
            \rho_{\Phi^-}=\left(
          \begin{array}{cccc}
                +0.48&0&0&-0.44\\
                0&+0.00&0&0\\
                0.00&0&+0.00&0.00\\
                -0.44&0&0.00&+0.52
                \end{array}
                \right)~.
            \end{equation}

Off-diagonal elements at detection time: $r_{18}^{\Phi^-}=r_{81}^{\Phi^-}=-0.33\pm0.04$.
Quoted values were reconstructed at preparation time due to the measured decoherence time $T_{\Phi^-}=210\pm10\,\mbox{ns.}$

    \begin{equation}
            \rho_{\Phi^+}=\left(
            \begin{array}{cccc}
               -0.48&0&0.00&+0.45\\
                0&+0.00&0&0\\
                0.00&0&+0.01&0.00\\
                +0.45&0&0.00&+0.52
                \end{array}
                \right)\,.
            \end{equation}

Off-diagonal elements at detection time: $r_{18}^{\Phi^+}=r_{81}^{\Phi^+}=0.34\pm0.04$.
Quoted values were reconstructed at preparation time due to the measured decoherence time $T_{\Phi^-}=213\pm10\,\mbox{ns.}$

In order to quantify the accuracy of the experimental data with respect to the theoretical expectations we define the fidelity as a mean square deviation as
\begin{equation}\label{eqn:fidelity}
    F_\rho = 1 - \frac{Tr\{(\rho_{exp}-\rho_{th})^2\}}{Tr\{\rho_{th}^2\}}
\end{equation}

which equals 1 in the ideal case and obeys the relation $0\leq F_\rho\leq 1$. The fidelities of all four $8\times8$ density matrices corresponding to the four Bell states are summarized in Table I.

\begin{table}[hbt]
        \begin{center}
            \begin{tabular}{c|c|c|c|c}
            & $\Psi^-$ & $\Psi^+$ & $\Phi^+$ & $\Phi^-$\\\hline
            $F_{\mathrm{P}\rho}$ & 0.97 & 0.97 & 0.98 & 0.98
            \end{tabular}
        \end{center}
        \caption{Calculated fidelities of the experimentally determined $8\times8$ density matrices $F_{\mathrm{P}\rho}$ according to Eq. \eqref{eqn:fidelity}.}
\end{table}

\section{Outlook}
The preparation of the entangled states presented here is based
on the pseudo pure concept of ensemble NMR quantum computing. It
has been shown theoretically that the corresponding mixed state
density matrices are separable and do not represent quantum
entanglement in the strictest sense \cite{braunstein:99,
schack:99}. Even if we would deal with a single $^{15}N@C_{60}$
molecule the pseudo entangled states presented here would be
mixed states at the temperature of $T = 50~K$ and microwave
frequency $\nu = 9.5~GHz$ applied here. However, in case we would
start out from a reasonably pure state and perform the same
preparation and tomography scenario as presented here we would
indeed reach a quantum entangled state. No other unitary
transformations as presented here would be required and in fact
the same signatures of the tomography like the characteristic
phase dependence would be observed. In this sense we consider the
experiments presented here as precursors of the corresponding
quantum experiments. In the following we try to estimate which
experimental parameters are required to reach such a quantum
state. We start from the Boltzmann density matrix according to
Eq.(\ref{eqn:rhoB}) where the inverse temperature $\beta =
\hbar/(k_\mathrm{B}T)$ and the Larmor frequency of the electron
spin $\omega_S = 2\pi\nu_S$ play the dominant role. We neglect
any Boltzmann polarization of the nuclear spins because of their
low Larmor frequency. In order to produce a spin alignment as
initial state we first apply a $\beta_0 = \pi$ pulse at the
electron spin transitions with $m_I = -1/2$. On this initial
state we apply the unitary sequence according to Eq.
(\ref{psiprep}). This leads to a density matrix which supposedly
represents a quantum entangled state for certain values of
temperature T and frequency $\omega_S$. All eight eigenvalues of
$\rho_B$ are positive at any frequency $\omega_S$ and temperature
$T$.

In order to apply the positive partial transpose (PPT) criterion
of Peres and Horodecki \cite{peres:96,horodecki:96} if a mixed
state density matrix is separable or not we need to perform the
partial transpose on one of the spins. If all eigenvalues are
still positive after partial transpose, the density matrix is
separable. If at least one of the eigenvalues becomes negative
there is some degree of entanglement. After performing the
partial transpose on the nuclear spin $I$ we obtain a zero
crossing of one of the eigenvalues under the condition

\begin{equation}\label{eqn:ebetaomega}
    e^{\beta\omega_S} \geq \left(3+2\sqrt{2}\right)^{\frac{1}{3}}\mbox { or }
    \beta\omega_S \geq \frac{1}{3}\log\left(3+2\sqrt{2}\right).
\end{equation}

The value for the quantum critical temperature below which the quantum regime is reached can be expressed as

\begin{equation}\label{eqn:Tq}
    T_q = \frac{\hbar\omega_S}{k_B\frac{1}{3}\log\left(3+2\sqrt{2}\right)}.
\end{equation}

For a high field ESR spectrometer operating at 95 GHz one obtains
$T_q = 7.73~K$. This is well in reach of current experimental
setups. It demonstrates that without the usual dynamical nuclear
polarization (DNP)of nuclear spins quantum states can be created
and correspondingly quantum algorithms can be performed.
Experiments along these lines are planned.

\section{Summary}

We have shown in detail experimental results on how to prepare
pseudo entangled states in a confined spin system, namely
$^{15}N@C_{60}$ consisting of an electron spin $S=3/2$ and
nuclear spin $I=1/2$. All four Bell states of a two qubit system
were produced. Moreover, we have performed  several variants of
spin density matrix tomography in order to verify the degree of
entanglement. By estimating the quantum critical temperature $T_q
= 7.73~K$ for an ESR frequency of $95~GHz$ we could show that
true quantum states are in reach with current technology.

\begin{acknowledgments}
    We acknowledge financial support by the German
    Bundesministerium f\"ur Bildung und Forschung (BMBF) and the Landestiftung Baden W\"urttemberg.
\end{acknowledgments}

\bibliographystyle{prsty}

\end{document}